\documentclass[%
aps,
superscriptaddress,
prl,%
preprint,%
reprint,%
floatfix
]{revtex4-1}

\usepackage{amsmath}
\usepackage{comment}
\usepackage{dcolumn}

\usepackage{natbib}
\usepackage{chngcntr}

\usepackage[colorlinks=true, allcolors=blue]{hyperref}

\usepackage{filecontents}

\usepackage{placeins}
\usepackage{color}
\usepackage{graphicx}
\usepackage{epstopdf}

\newcommand{\pten}[2]{${#1\times10^{#2}}$}

\usepackage[caption=false]{subfig}

\newcommand{\umob}{cm$^2$V$^{-1}$s$^{-1}$}

\newcommand{\kv}{{\bf k}}
\newcommand{\qv}{{\bf q}}

\begin{document}

\author{A. V. Lugovskoi}
\affiliation{\mbox{Institute for Molecules and Materials, Radboud University, Heijendaalseweg 135, NL-6525 AJ Nijmegen, Netherlands}}
\author{M. I. Katsnelson}
\affiliation{\mbox{Institute for Molecules and Materials, Radboud University, Heijendaalseweg 135, NL-6525 AJ Nijmegen, Netherlands}}
\affiliation{\mbox{Theoretical Physics and Applied Mathematics Department,
Ural Federal University, 620002 Ekaterinburg, Russia}}
\author{A. N. Rudenko}
\email{a.rudenko@science.ru.nl}
\affiliation{School of Physics and Technology, Wuhan University, Wuhan 430072, China}
\affiliation{\mbox{Institute for Molecules and Materials, Radboud University, Heijendaalseweg 135, NL-6525 AJ Nijmegen, Netherlands}}
\affiliation{\mbox{Theoretical Physics and Applied Mathematics Department,
Ural Federal University, 620002 Ekaterinburg, Russia}}

\date{\today}
	
\title{
Strong electron-phonon coupling and its influence on the transport \\  and optical properties of hole-doped single-layer InSe
}

\begin{abstract} 
 We show that hole states in recently discovered single-layer InSe are strongly renormalized by the coupling with acoustic phonons. The coupling is enhanced significantly at moderate hole doping ($\sim$10$^{13}$ cm$^{-2}$) due to hexagonal warping of the Fermi
 surface. While the system remains dynamically stable, its electron-phonon spectral function exhibits sharp low-energy resonances, leading to the formation of satellite quasiparticle states near the Fermi energy. Such many-body renormalization is predicted to have two important consequences. First, it significantly suppresses charge carrier mobility reaching $\sim$1 cm$^2$V$^{-1}$s$^{-1}$ at $100$ K in a freestanding sample. Second, it gives rise to unusual temperature-dependent optical excitations in the midinfrared region. Relatively small charge carrier concentrations and realistic
 temperatures suggest that these excitations may be observed experimentally.
\end{abstract}

\maketitle

{\it Introduction.}---Two-dimensional (2D) indium selenide (InSe) is a recently discovered semiconductor receiving considerable attention because of its attractive electronic properties. Thin films of InSe have been proposed to be suitable for field-effect transistor applications due to their high carrier mobilities, reported to exceed 10$^3$ \umob~at room temperature \cite{Sucharitakul_2015, Bandurin2017, Ho_2017, Lin2018}.
Other interesting properties of this material include a tunable band gap \cite{Mudd2013,Lei2014,Brotons-Gisbert_2016,Magorrian2016}, fully developed quantum Hall effect~\cite{Bandurin2017}, anomalous optical response~\cite{Bandurin2017,Mudd2013,Tamalampudi_2014,Wei_2015}, superior flexibility \cite{Zhao2019}, as well as excellent thermal \cite{Nissimagoudar_2017,Zhou_2018} and thermoelectric \cite{Hung2017} characteristics. These observations, along with its ambient stability \cite{Politano2016}, make low-dimensional InSe an appealing candidate for numerous practical applications \cite{Boukhvalov2017}.

Single-layer (SL) InSe is a layered semiconductor with an indirect energy gap in the visible range. Its  electronic structure is characterized by peculiar flat regions in the valence band~\cite{Zolyomi2014}, giving rise to a giant van Hove singularity in the hole density of states \cite{Rybkovskiy2014,Eriksson2015}. This feature is known as a ``Mexican-hat''-like band, and has been predicted for the whole family of In$_2X_2$ and Ga$_2X_2$ ($X$=S,Se,Te) compounds
~\cite{Zolyomi2014,Robertson_1979}. Interestingly, this peculiar shape evolves into the conventional parabolic band with increasing material's thickness, as has been theoretically predicted for InSe~\cite{Rybkovskiy2014,Mudd2016} and recently confirmed by angular resolved photoemission spectroscopy~\cite{Kibirev}. This observation makes ultrathin InSe films especially attractive for further studies.
The interest to flat-band materials is motivated by exotic physical properties of such systems, closely related to various many-body instabilities and strong correlation effects~\cite{Khodel1990,Volovik1991,Nozieres1992,Irkhin2002,Yudin2014}. Recent experimental discovery of flat bands and associated many-body phenomena including superconductivity in magic-angle twisted bilayer graphene~\cite{Cao2018a,Cao2018b} enhances enormously the interest to the problem.

Electron-phonon interaction in the case of narrow bands can be dramatically different from the conventional case and can play a more important role. In particular, a strong electron-phonon coupling in the magic-angle twisted bilayer graphene was recently reported~\cite{Choi2018}. When typical electron energies are comparable with the phonon ones, one can expect essential nonadiabatic modifications of electron~\cite{Hewson1981} and phonon~\cite{Katsnelson1985} energy spectra. For the case of a single electron in a crystal we have a well-known and well-studied problem of polaron~\cite{FrohlichAdvPhys54,FeynmanPR55,AppelPolarons,EminPolarons}. The situation when we have a {\it degenerate} electron gas with a small electron energy scale near the Fermi energy is still quite poorly investigated.

In this work, we use many-body theory combined with first-principles calculations to study electron-phonon coupling and related properties of hole-doped SL-InSe. We find that due to peculiar character of hole states in SL-InSe, their interaction with phonons is anomalously strong. We predict that this interaction leads to a considerable limitation of charge carrier mobility in freestanding SL-InSe samples, but gives rise to unexpected temperature-dependent midinfrared optical excitations with large spectral weight.

{\it Structure and basic electronic properties of SL-InSe.}---Depending on the chemical composition, In-Se binary compounds demonstrate a rich variety of ordered phases with different physical properties~\cite{Han_2014}. Among structural forms with a 1:1 stoichiometry, the $\gamma$ polytype is the most attractive phase for exfoliation, composed of vertically stacked weakly interacting InSe layers. Each layer adopts a crystal structure with two vertically displaced 2D buckled honeycombs ($D_{3h}$ point group) \cite{SI}.
Atomically thin InSe was experimentally obtained by means of mechanical exfoliation~\cite{Mudd2013,Brotons-Gisbert_2016,Bandurin2017,Lei2014}, as well as by epitaxial growth from a liquid phase~\cite{Lauth_2016}.

\begin{figure}[b]
	\centering
	\includegraphics[width=0.5\textwidth]{{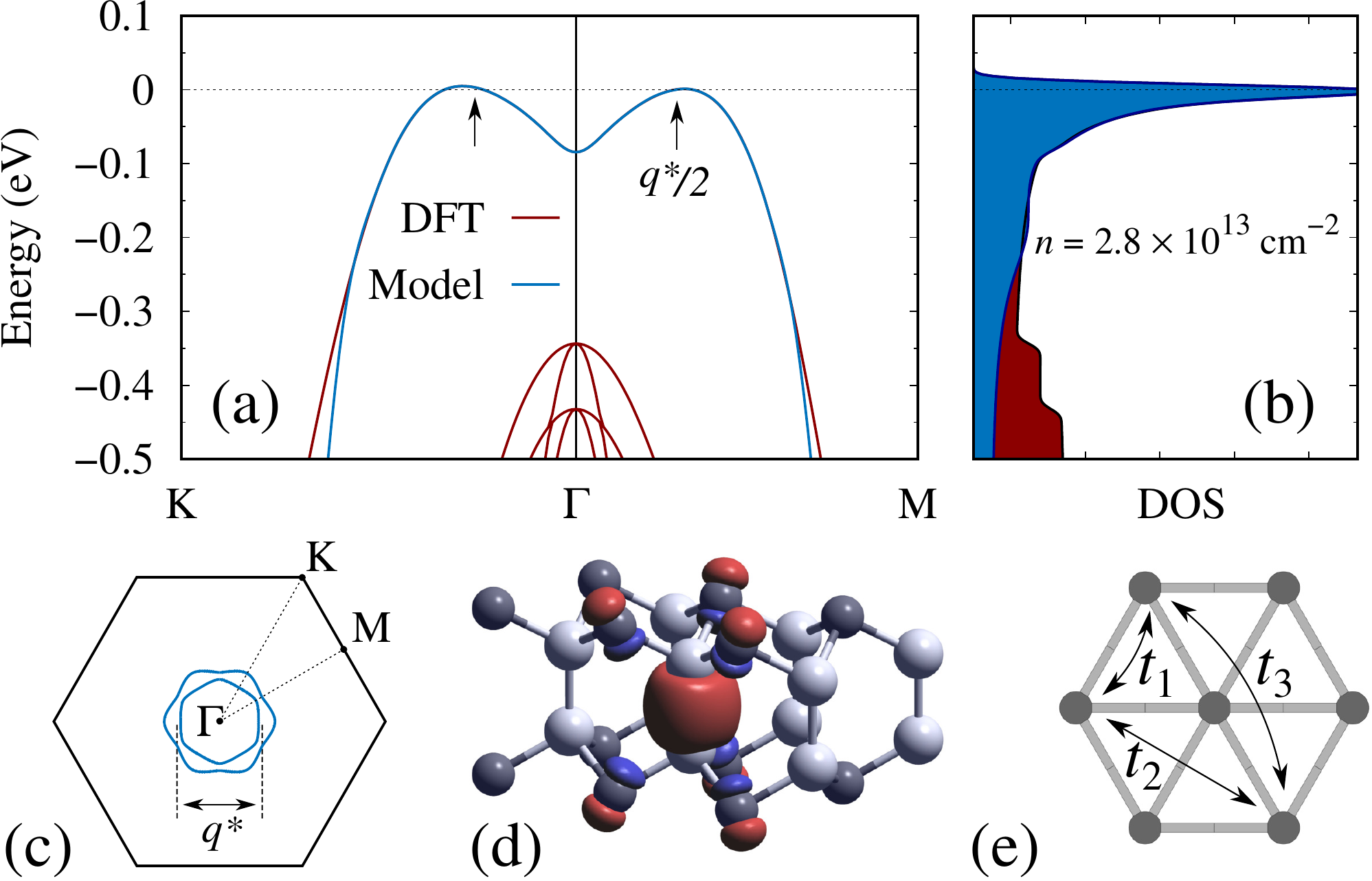}}
	\caption{(a) Band structure and (b) DOS of hole-doped SL-InSe for the concentration $n=2.8\times10^{13}$ cm$^{-2}$ calculated using DFT and TB model as described in the text. (c) Brillouin zone and the Fermi surface with $q^{*}$ denoting the characteristic scattering wave vector. (d) Real-space wave function of the hole state, where light and dark spheres denote In and Se atoms, respectively. (e) Effective lattice model for holes in SL-InSe with $t_1$, $t_2$, $t_3$ being the dominant interaction parameters.}
	\label{fig1}
\end{figure}

Density-functional theory (DFT) electronic structure calculations and structural optimization were performed in this work by means of the plane-wave {\sc quantum espresso} ({\sc qe}) code~\cite{Giannozzi2017},
using fully relativistic norm-conserving pseudopotentials \cite{PP}.
Exchange and correlation were treated with the local density approximation.
The kinetic energy cutoff for plane waves was set to 80 Ry, and the Brillouin zone (BZ) was sampled with a (80$\times$80) \kv-point mesh
to ensure numerical accuracy. A vacuum thickness of 30 \AA~was introduced to avoid spurious interactions between the periodic supercell images in the direction perpendicular to the 2D plane.
Carrier doping was introduced at the DFT level by increasing the total charge of the system and adding the compensating jellium background to preserve the charge neutrality.

In Fig.~\ref{fig1}(a), we show the valence band of SL-InSe calculated for the hole concentration of $n=2.8 \times 10^{13}$ cm$^{-2}$. The prominent characteristic of hole doped SL-InSe is the van Hove singularity in the density of states (DOS) near the Fermi energy [Fig.~\ref{fig1}(b)] originating from the hexagonal warping of the Fermi surface [Fig.~\ref{fig1}(c)]. The strong nesting of the Fermi surface is expected to be responsible for giant anomalies in the response functions at the characteristic wave vector $q^* \approx 0.56$ \AA$^{-1}$. In the energy region up to 0.3 eV, the valence states are represented by a single ``Mexican hat''-like band. Importantly, this peculiar shape is only typical for SL- and few-layer InSe, while evolving into the parabolic shape for thick samples \cite{Kibirev}. For the purpose of our study, it is convenient to map the full DFT Hamiltonian onto a single-band tight-binding (TB) model. To this end, we use the formalism of maximally localized Wannier functions (WF) \cite{Marzari2012} as implemented in {\sc wannier90} code \cite{wannier90}. The resulting model is defined in the basis of WFs localized in the center of an In-In bond with tails on neighboring Se atoms, as shown in Fig.~\ref{fig1}(d). The effective TB Hamiltonian is defined on a triangular lattice [Fig.~\ref{fig1}(e)] and is determined by the leading hopping integrals: $t_1=0.24$ eV, $t_2=-0.10$ eV, and $t_3=-0.06$ eV. In the low-energy region the model perfectly matches the first-principles results.

\begin{figure}[b]
	\centering
	\includegraphics[width=0.52\textwidth]{{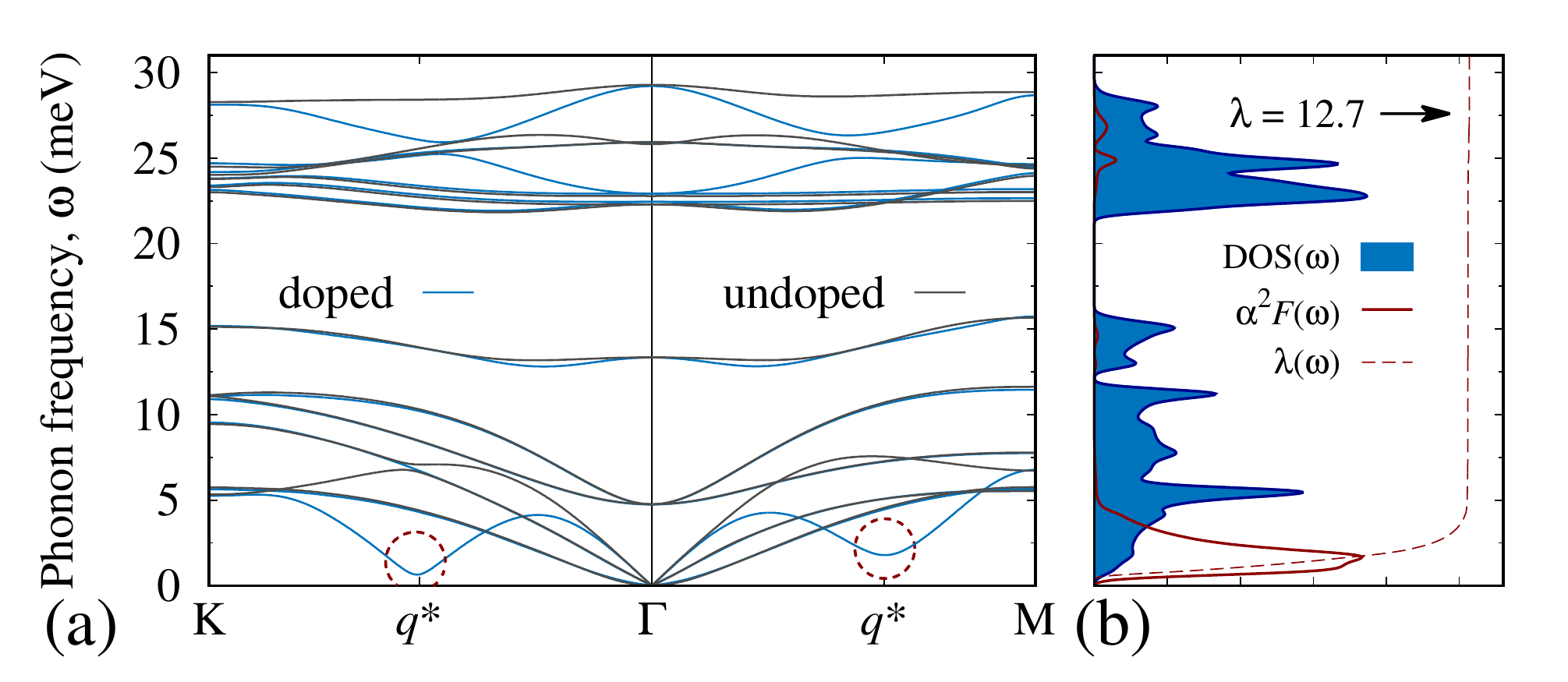}}
	\mbox{
	\includegraphics[width=0.255\textwidth]{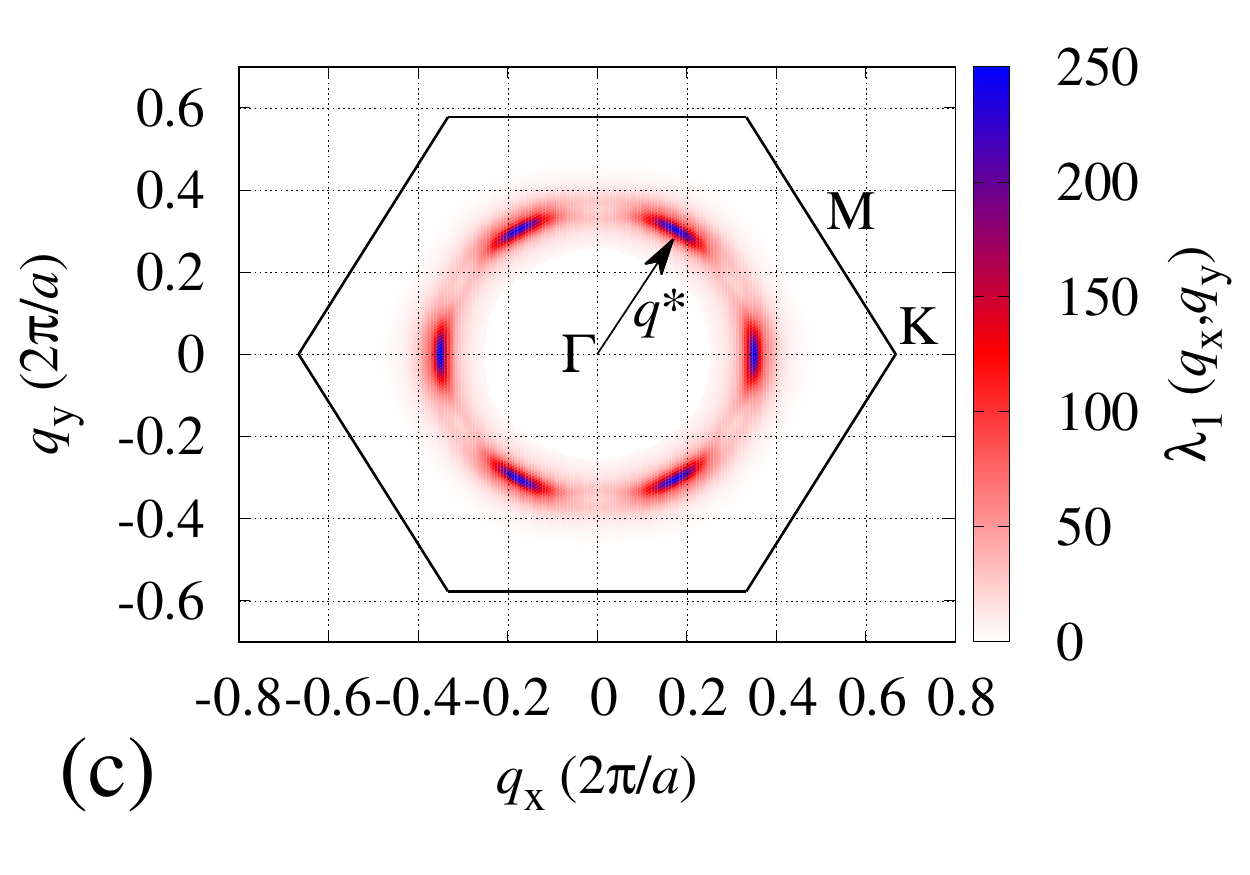}
	\includegraphics[width=0.255\textwidth]{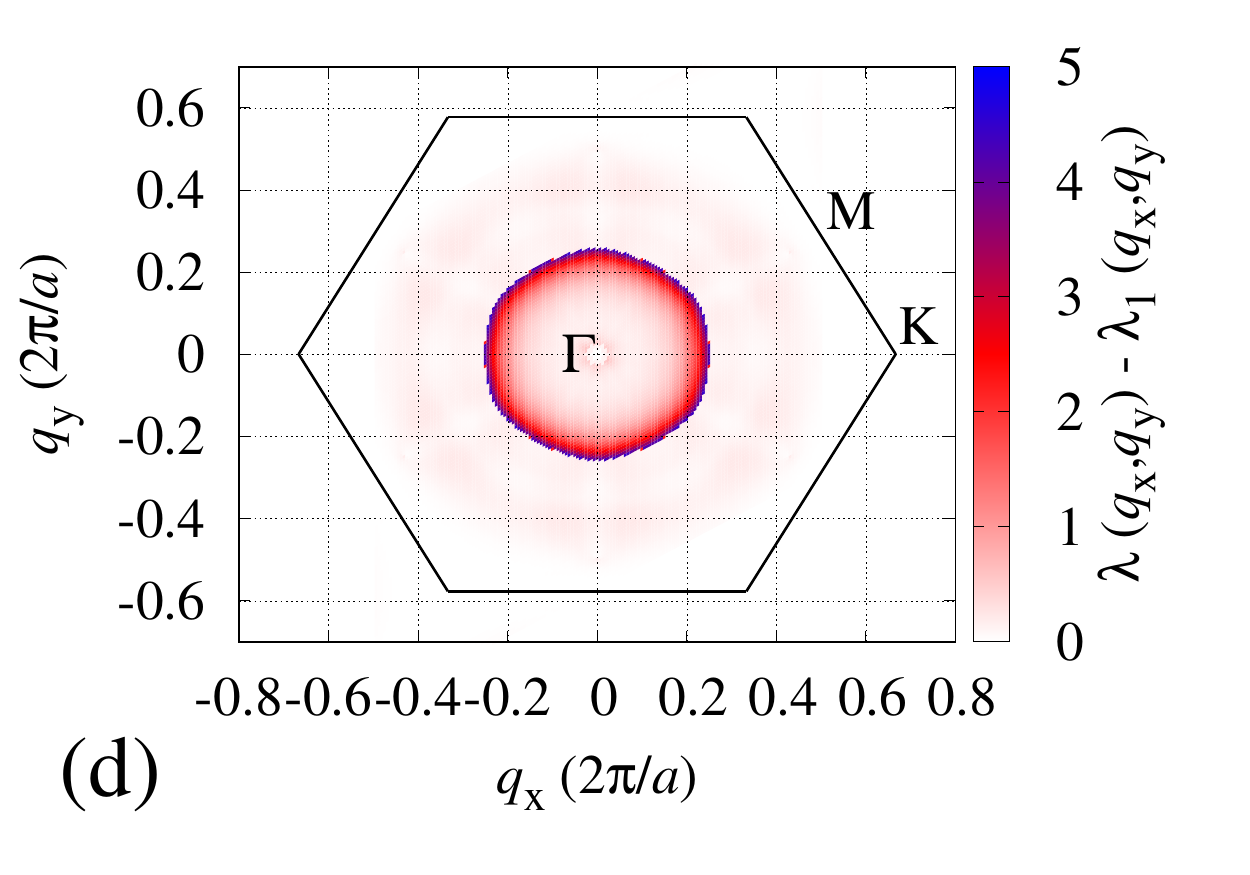}
	}
	\caption{(a) Phonon dispersion of undoped and hole-doped SL-InSe calculated for the concentration $n=2.8\times10^{13}$ cm$^{-2}$.  Circles show softening of the phonon mode at the characteristic wave vector $q^*$. (b) Frequency-dependent phonon DOS, electron-phonon spectral function $\alpha^2 F(\omega)$, and integrated electron-phonon coupling strength $\lambda(\omega)=2\int\,d\omega\,\omega^{-1} \alpha^2 F(\omega)$. (c),(d) Momentum-resolved electron-phonon coupling for the lowermost acoustic branch ($\lambda_1$) and for all other branches, respectively.}
	\label{fig2}
\end{figure}

{\it Phonons and electron-phonon coupling.}---The phonon-related properties are calculated using density functional perturbation theory (DFPT) \cite{BaroniRMP,Giustino2017} as implemented in {\sc qe}. The calculations are performed on a (16$\times$16) \qv-point mesh.
The phonon dispersion curves and the corresponding DOS are shown in Figs.~\ref{fig2}(a) and \ref{fig2}(b) for undoped and hole-doped SL-InSe. The undoped spectrum is typical to 2D materials, with the prominent out-of-plane (ZA) mode having quadratic dispersion around the $\Gamma$ point. Unlike elemental 2D semiconductors \cite{Zhu2014,Lugovskoi}, optical modes in SL-InSe appear already at 5 meV, giving rise to a sharp peak in DOS. The hole-doping leads to a softening of the lowermost acoustic phonon mode at wave vector $q^{*}$, which corresponds to the nesting wave vector depicted in Fig.~\ref{fig1}(c). The frequencies of the soft mode depend on the doping value remaining positive up to the critical doping $n^*\approx 2.9 \times 10^{13}$ cm$^{-2}$, at which imaginary frequencies appear suggesting structural instability of heavily doped samples. The rest of the phonon spectrum remains virtually unaffected by moderate hole doping.

We now turn to the interaction of phonons with holes in SL-InSe.
To the first order in atomic displacements, the interaction of electrons with phonons can be described by the vertex $g^{\nu}_{\kv \kv'}=l_{\nu\qv}M^{\nu}_{\kv \kv'}$, where $l_{\nu\qv}=\sqrt{\hbar/2m_0\omega_{\nu\qv}}$ is the characteristic displacement, and $M^{\nu}_{\kv \kv'}=\langle \psi_{\kv'}|\partial_{-\qv\nu} V| \psi_{\kv} \rangle$ is the matrix element of the interaction potential $\partial_{-\qv\nu} V$ associated with electronic states $\psi_{\kv'}$ and $\psi_{\kv}$, phonon wave vector $\qv=\kv'-\kv$, branch index $\nu$, and frequency $\omega_{\nu\qv}$. In order to reach numerical convergence in evaluating BZ zone integrals, the electron-phonon-related quantities have been interpolated in momentum space by making use of the Wannier-interpolation scheme \cite{Giustino2007}. To this end, we utilize the {\sc epw} package \cite{Ponce2016}, which allows us to achieve sufficiently accurate BZ sampling using (160$\times$160) \kv- and (320$\times$320) \qv-point meshes.

To quantify the electron-phonon coupling strength in metallic systems, it is convenient to define a dimensionless quantity averaged over the Fermi surface as $\lambda_{\qv \nu}=\langle |g^{\nu}_{\qv}|^2 \rangle /\omega_{\qv \nu}$, also known as the mass enhancement factor. The spectral representation of this quantity is known as the Eliashberg spectral function, which reads
\begin{equation}
\alpha^{2}F( \omega) = \frac{1}{2} \sum_{{\bf q}\nu}
 \omega_{{\bf q}\nu} \lambda_{{\bf q}\nu}
\delta (\omega - \omega_{{\bf q}\nu})
.
\label{eq:a2f}
\end{equation}
In Figs.~\ref{fig2}(c) and \ref{fig2}(d), we show $\lambda_{\qv \nu}$ calculated for the lowermost (soft) phonon mode ($\nu=1$), as well as for all other modes in hole-doped SL-InSe. The dominant electron-phonon coupling originates from the interaction with the soft phonon mode, exceeding the contribution from all other modes by almost 2 orders of magnitude. As expected, the strongest coupling is taking place around the nesting wave vector $q^*$. In the frequency domain, the coupling is localized at low frequencies, giving rise to a pronounced peak below 5 meV in the Eliashberg function shown in Fig.~\ref{fig2}(b). The total interaction $\lambda=\sum_{\qv\nu}\lambda_{\qv\nu}$ is doping dependent and reaches its maximum value of $\sim $13 around the critical value $n^*$ (see Table \ref{table1}). In all cases considered, $\lambda \gg 1$, which allows us to designate hole-doped SL-InSe as a system with strong electron-phonon coupling, and expect significant renormalization of the observable electronic properties.

{\it Many-body renormalization.}---In the interacting system, the electronic structure can be characterized by the renormalized finite-temperature Green's function $G^{-1}_{\kv}(i\omega_j)=G^{-1}_{0\kv}(i\omega_j)+\Sigma_{\kv}(i\omega_j)$, where $G_{0\kv}(i\omega_j)=(i\omega_j+\mu+\varepsilon_{\kv})^{-1}$ is the bare (noninteracting) Green's function, and $\Sigma_{\kv}(i\omega_j)$ is the electronic self-energy, describing the interaction. According to the Migdal theorem \cite{Migdal1958}, the latter can be approximated by a diagram involving electron $G_{\kv}(i\omega_j)$ and phonon $D_{\qv\nu}(i\omega_{j'})$ propagators connected by the electron-phonon vertices $g^{\nu}_{\kv\kv'}$, i.e.
\begin{equation}
    \Sigma_{\kv}(i\omega_j)=-T\sum_{\kv'\nu j'}G_{\kv}(i\omega_j)|g^{\nu}_{\kv\kv'}|^2 D_{\kv-\kv'\nu}(i\omega_j-i\omega_{j'}),
\end{equation}
where $i\omega_j=i(2j+1)\pi T$ are the fermionic Matsubara frequencies with $j$ being an integer number \cite{SI}.
This approximation assumes neglecting vertex corrections in the self-energy expansion~\cite{Migdal1958}. Generally speaking, its applicability for the case of narrow energy bands is not fully justified. However, given that any realistic calculations of electron-phonon interaction with vertex corrections are far beyond current computational capabilities, we have no alternatives. Nevertheless, we believe that the Migdal approximation is at least qualitatively correct in our case.

\begin{figure}[tbp]
	\centering
	\includegraphics[width=0.45\textwidth]{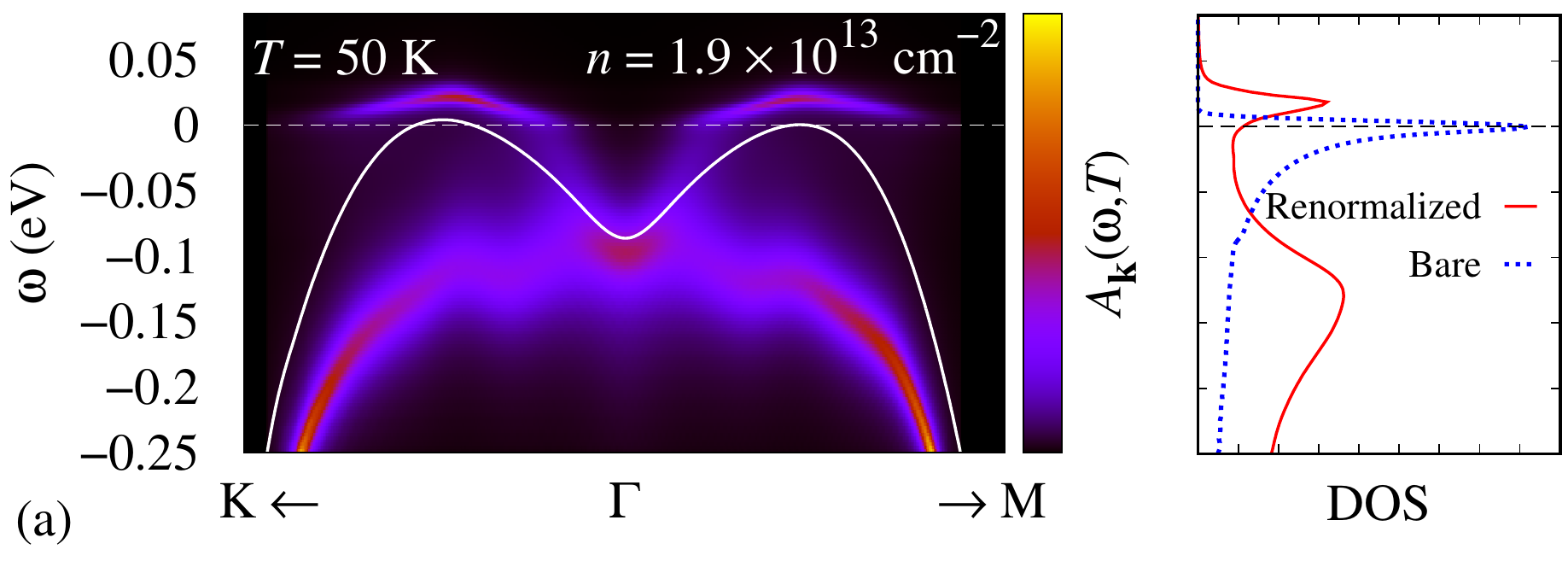}
	\includegraphics[width=0.45\textwidth]{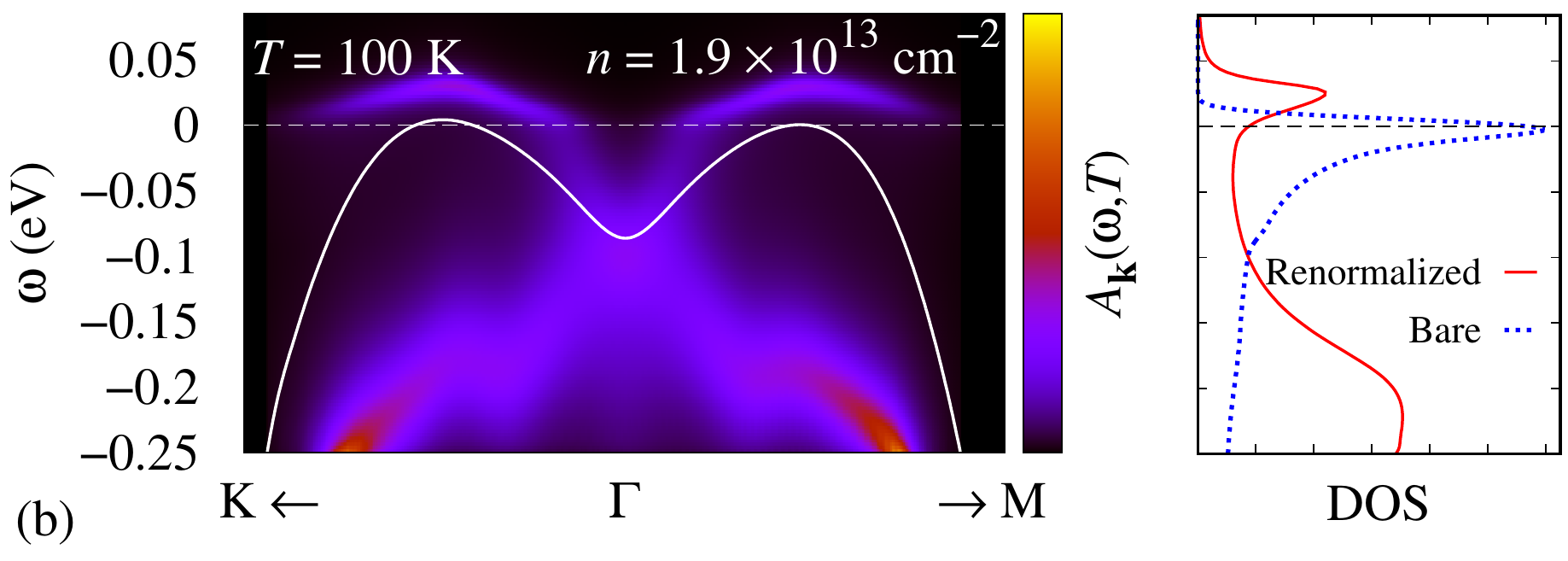}
	\includegraphics[width=0.45\textwidth]{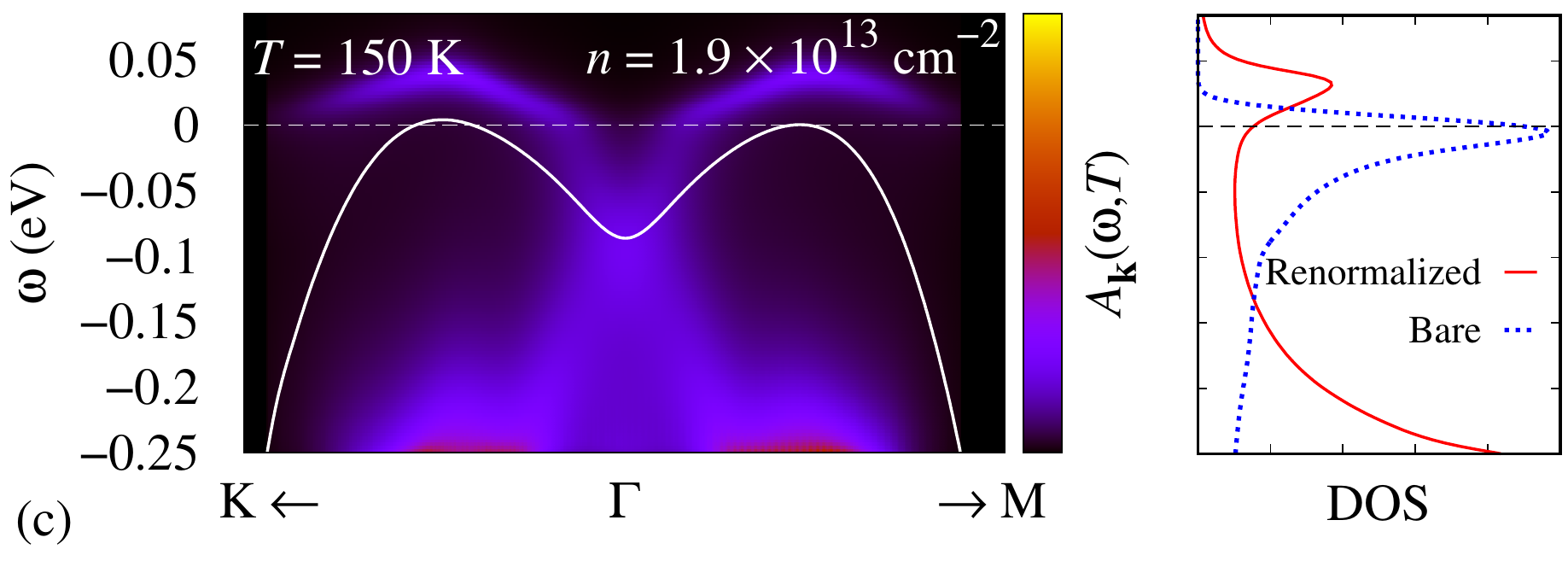}
	\caption{Energy- and momentum-resolved spectral function $A_{\kv}(\omega,T)$ and corresponding density of states $\sum_{\kv}A_{\kv}(\omega,T)$ calculated for hole-doped SL-InSe in the presence of electron-phonon coupling at carrier concentration $n=1.9\times10^{13}$ cm$^{-2}$ for three different temperatures: (a) 50 K, (b) 100 K, and (c) 150 K. White solid and blue dashed curves correspond to the noninteracting (bare) bands and DOS, respectively. For clarity, the renormalized DOS (red solid curve) is enlarged by a factor of 2.}
	\label{fig3}
\end{figure}

In Fig.~\ref{fig3}, we show the spectral function $A_{\kv}(\omega,T)=-\frac{1}{\pi}\mathrm{Im}[G_\kv(\omega)]$ calculated in SL-InSe for $T=$ 50, 100, and 150 K at hole concentration $n=1.9\times 10^{13}$ cm$^{-2}$. The hole spectrum exhibits a significant renormalization compared to the noninteracting one. The spectrum reconstruction is taking place in the vicinity of the characteristic wave vector $q^*/2$ at which the hole states asymmetrically split into the upper and lower subbands. The splitting originates
from the discontinuity in the self-energy $\Sigma_{\kv}(\omega,T)$ at $|\kv| \sim q^*/2$ due to the strong electron-phonon coupling. The self-energy is dependent on the phonon occupation $n_{\qv\nu}(T)$, resulting in the temperature dependence of the observed splitting. Moreover, since around $T=100$ K acoustic phonons can be considered classically ($n_{\qv\nu} \simeq T/\omega_{\qv\nu}$), this dependence is nearly linear. Both subbands are clearly observed in temperature-dependent DOS [Figs.~\ref{fig3}(a)--(c)], which allows one to expect optically active transitions between them.

\begin{table}[tbp]
        \caption{Dimensionless electron-phonon coupling constant $\lambda$, phonon-limited carrier mobility $\mu$ (at $T=100$ K), and characteristic optical frequency $\omega^*$ corresponding to the local maximum of the conductivity $\sigma(\omega^*)$ (at $T=100$ K) shown at different values of hole doping ($n$) in SL-InSe.}
        \begin{tabular*}{1.0\linewidth}{@{\extracolsep{\fill}}ccccc}
                \hline \hline
                $n$ (cm$^{-2}$) &  \quad $\lambda$ \quad & $\mu$ (cm$^{2}$V$^{-1}$s$^{-1}$) & \quad $\omega^*$ (eV) \quad & $\sigma(\omega^*)/\sigma(0)$ \\ \hline
                \pten{8.17}{12} & \quad 4.6 \quad &  1.92 & \quad 0.26 \quad & 0.58 \\
                \pten{1.93}{13} & \quad 5.9 \quad &  0.94 & \quad 0.28 \quad & 0.73 \\
                \pten{2.82}{13} & \quad 12.7 \quad & 0.40 & \quad 0.37 \quad & 1.22 \\
                \hline\hline
        \end{tabular*}
        \label{table1}
\end{table}

{\it Transport and optical properties.}---Having calculated the interacting Green's function, as a next step we consider transport and optical properties of the system. Within the linear-response formalism, the conductivity tensor can be written as \cite{Green-Book}
\begin{equation}
    \mathrm{Re}[\sigma_{\alpha\beta}(\qv,\omega)] = -\frac{1}{\omega}\mathrm{Im}[\chi_{\alpha\beta}(\qv,\omega)],
\end{equation}
where $\chi_{\alpha\beta}(\qv,\omega)=\langle\langle j_{\alpha}(\qv,\omega)j_{\beta}(-\qv,\omega)\rangle\rangle$ is the current-current correlation function with $\langle \langle ... \rangle \rangle$ being the Kubo correlation function. Ignoring the vertex corrections, the corresponding quantity can be written in the limit of zero momentum transfer ($\qv \rightarrow 0$) as
\begin{equation}
    \chi_{\alpha\beta}(0,i\omega_j) = e^2T \sum_{\kv \omega_j'} v^{\alpha}_{\kv} G_{\kv}(i\omega_j') v^{\beta}_{\kv} G_{\kv}(i\omega_j'+i\omega_j),
\end{equation}
where $v^{\alpha}_{\kv}=\hbar^{-1}\partial \varepsilon_{\kv}/\partial k_{\alpha}$ is the $\alpha$ component of the group velocity.
In what follows, we consider diagonal components of the conductivity tensor only. Moreover, taking into account isotropic behavior of the system at $\qv \rightarrow 0$, we have $\sigma\equiv \sigma_{xx}=\sigma_{yy}$.

\begin{figure}[tbp]
	\centering
	\includegraphics[width=0.4\textwidth]{{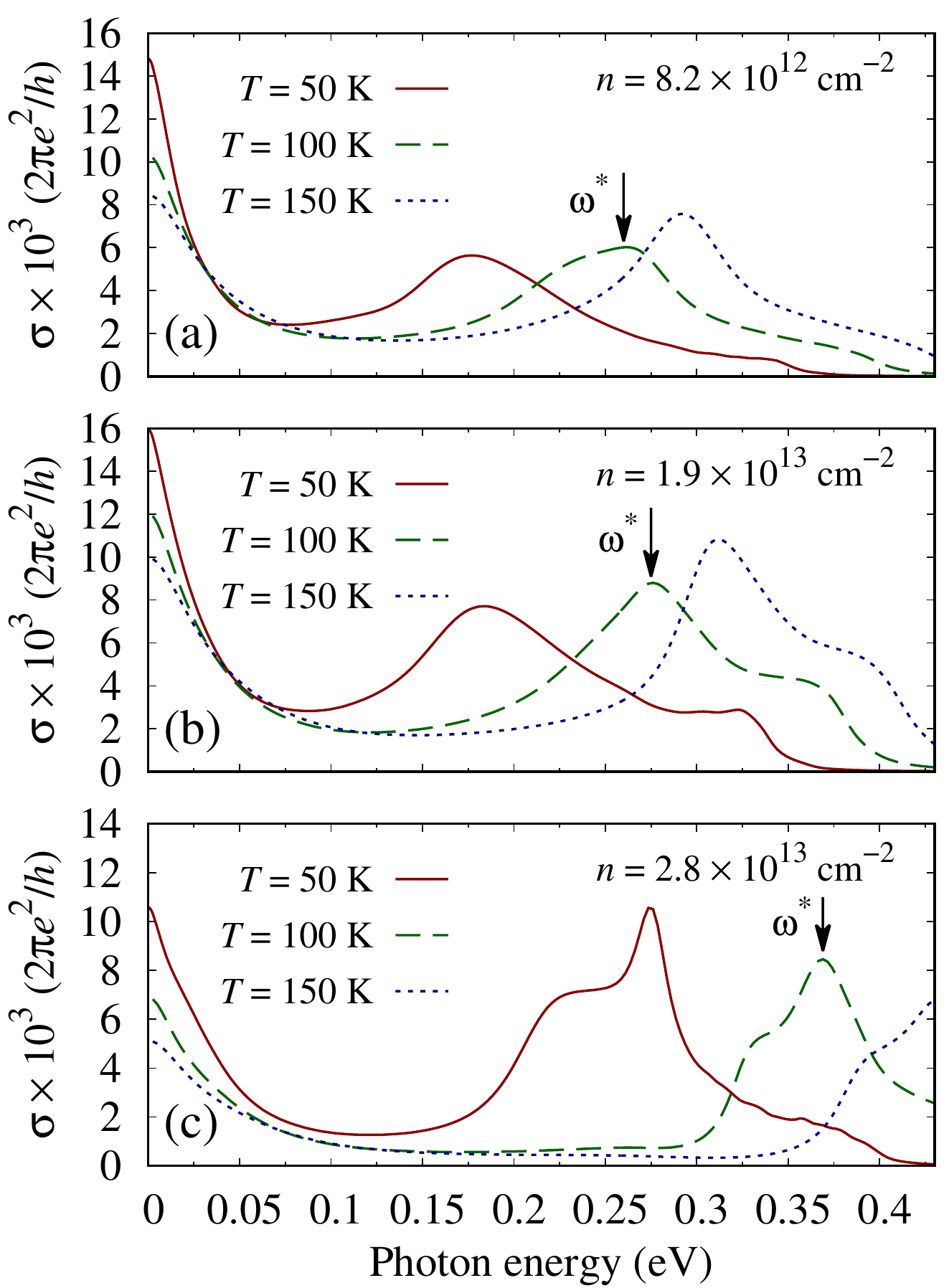}}
	\caption{Temperature-dependent optical conductivity in SL-InSe calculated in the presence of electron-phonon coupling for different values of hole doping. $\omega^*$ is the characteristic optical frequency listed in Table \ref{table1} with the corresponding conductivity values.}
	\label{fig4}
\end{figure}

In Fig.~\ref{fig4}, we show optical conductivity calculated in SL-InSe as a function of the photon energy for different temperatures and hole concentrations.
At low energies, typical metallic behavior is observed with a clear Drude peak, whose amplitude in our case has the meaning of the phonon-limited direct current (dc) conductivity, i.e. $\sigma^{\mathrm{dc}}\equiv\sigma(\omega \rightarrow 0)$. As temperature raises, dc conductivity decreases linearly with the temperature, demonstrating a dependence typical for metals in our temperature range. With increasing charge carrier concentration, dc conductivity tend to decrease. The relation of dc conductivity to the charge doping is more involved because of a nonlinear dependence of the electron-phonon coupling.
The corresponding hole mobilities $\mu=\sigma^{\mathrm{dc}}/ne$ calculated at $T=100$ K for different hole concentrations $n$ are listed in Table \ref{table1}.  In all cases we obtain remarkably low values of the order of 1 cm$^{2}$V$^{-1}$s$^{-1}$. This result is highly unexpected in view of recent experimental reports on high mobility in 2D InSe \cite{Bandurin2017,Sucharitakul_2015, Ho_2017,Lin2018}. We note, however, that our finding is solely applicable to a freestanding InSe single layer. Electronic structure of multilayer InSe, as well as encapsulated or supported samples turns out to be modified considerably by the interlayer interactions \cite{Kibirev}, suppressing strong electron-phonon coupling. From the theoretical viewpoint low mobility values in SL-InSe are well understood and can be attributed to the formation of a coupling-induced pseudogap in the hole spectrum [see Fig.~\ref{fig3}], resulting in a small DOS at the Fermi energy.

Finally, let us discuss many-body effects in the optical response of SL-InSe. In the independent particle approximation, the optical conductivity of an electron gas is given by the Drude formula, $\sigma(\omega)=\frac{\gamma}{4\pi}\frac{\omega^2_p}{\omega^2 + \gamma^2}$, with $\omega_p$ being the plasma frequency, and $\gamma$ is a phenomenological damping. The Drude model describes monotonically decreasing conductivity as a function of energy unless interband transitions come into play. Since we deal with an effectively single-band system, no interband transitions are expected below at least 0.35 eV [see Fig.~\ref{fig1}(a)]. Nevertheless, the calculated optical conductivity of hole-doped SL-InSe presented in Fig.~\ref{fig4} is qualitatively different. Specifically, the spectrum exhibits notable peaks in the energy range 0.1--0.3 eV, which are beyond the Drude model. The corresponding magnitudes are comparable with the Drude peaks at $\omega \rightarrow 0$, and is increasing with doping. The observed excitations originate from the many-body splitting of hole states induced by the strong electron-phonon coupling, as discussed earlier. Importantly, the resonant frequencies are strongly temperature-dependent, which is in favor for their uncomplicated experimental detection. Relatively small doping required to observe the effect as well as favorable (midinfrared) spectral range makes us believe that this prominent optical characteristic of hole-doped SL-InSe might be observable by conventional optical, or even scanning probe techniques.

{\it Conclusion.}---We have shown that moderately hole-doped SL-InSe demonstrates strong electron-phonon coupling. The coupling arises from the Fermi surface nesting combined with the doping-induced softening of acoustic phonon modes. We have found two important consequences of this phenomenon: (i) Charge carrier mobility turns out to be essentially suppressed, being of the order of 1 cm$^2$V$^{-1}$s$^{-1}$. (ii) Electronic spectrum undergoes strong renormalization, giving rise to unconventional temperature-dependent optical excitations in the midinfrared region with large spectral weight. Both effects are predicted for realistic doping levels ($\sim$10$^{13}$ cm$^{-2}$) and temperatures around 100 K. We anticipate that our findings could be verified by standard experimental techniques, provided that sufficiently clean single-layer samples could be fabricated in a weakly interacting environment.

 Our findings can motivate further experimental and theoretical studies on low-dimensional InSe. Apart from being an obvious candidate for experimental verification of the unusual optical response, SL-InSe is a promising material for exploring other aspects of many-body physics, including superconductivity, $sp$ magnetism, electron-correlation effects, as well as other collective phenomena.

\FloatBarrier
\begin{acknowledgments}
This work is part of the research program ``Two-dimensional semiconductor crystals'' with project number 14TWOD01, which is (partly) financed by the Netherlands Organization for Scientific Research (NWO).
M.I.K. acknowledges support from FLAG-ERA JTC2017 Project GRANSPORT. A.N.R. acknowledges support from the Russian Science Foundation, Grant No. 17-72-20041.
\end{acknowledgments}

{\it Note added in the proof.}---Recently, we became aware of publications discussing the transport properties of 2D InSe at the single-particle theory level \cite{note1,note2,note3}, which
might be of interest to the readers. Particularly, similar values for the hole mobility were obtained in Ref.~\cite{note3}.

\nocite{apsrev41Control}
\bibliographystyle{apsrev4-1}
\bibliography{inse+sm}

\clearpage
\onecolumngrid
\begin{center}
\textbf{\large Supplemental Material: Strong electron-phonon coupling and its influence on the transport and optical properties of hole-doped single-layer InSe}
\end{center}
\twocolumngrid

\setcounter{equation}{0}
\setcounter{figure}{0}
\setcounter{table}{0}
\makeatletter
\renewcommand{\theequation}{S\arabic{equation}}
\renewcommand{\thefigure}{S\arabic{figure}}

\subsection{Equilibrium crystal structure of SL-InSe}

\begin{figure}
        \includegraphics[]{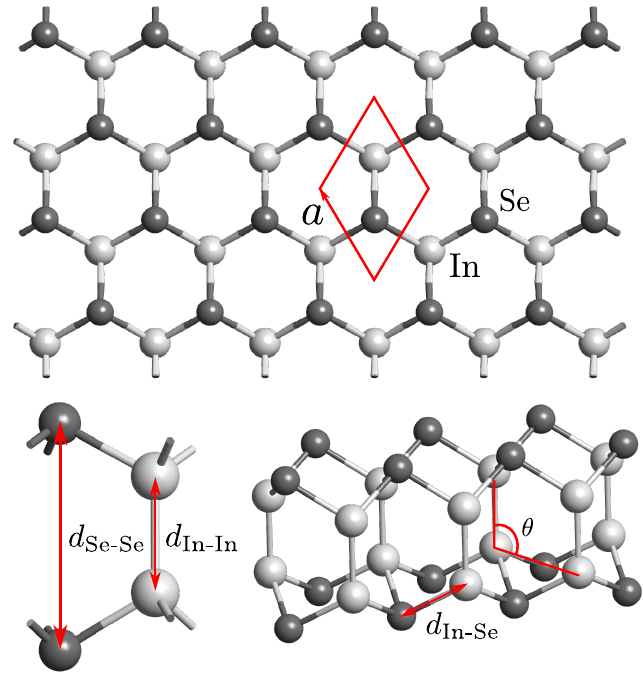}
        \caption{Schematic representations of the equilibrium crystal structure of SL-InSe considered in this work. Red lines denote the hexagonal unit cell and the following structural constants obtained from DFT calculations: $a=3.94$ \AA, $d_\text{Se-Se}=5.24$ \AA, $d_\text{In-In}=2.70$ \AA, $d_\text{In-Se}=2.61$ \AA,  and $\theta=90^{\text{o}}.$ }
        \label{fig:struct}
\end{figure}

Fig.~\ref{fig:struct} shows the equilibrium crystal structure of SL-InSe with the corresponding structural constants obtained from full structural optimization at the DFT level. The resulting values are in good agreement with the literature data \cite{Zolyomi2014}.

\subsection{Computational details}

Here, we provide some relevant expressions used in this work to calculate frequency-dependent characteristics related to the electron-phonon coupling. More details can be found, e.g., in Refs.~\onlinecite{Giustino2017,Green-Book}.

For a single-band system with first-order electron-phonon coupling, the electron self-energy in the Migdal approximation \cite{Migdal1958} can be represented on the real-frequency axis as:

\begin{widetext}
\begin{equation}
\Sigma_{\kv}(\omega,T)
=\sum_{\nu}\int_\text{BZ}\frac{d\qv}{\Omega_\text{BZ}}|g_{\nu}(\kv,\qv)|^{2}
\bigg[
\frac{n_{\qv\nu}(T)+f_{\kv+\qv}(T)}{\omega-(\varepsilon_{\kv+\qv}-\varepsilon_\text{F})+\omega_{\qv\nu}+i\eta}
+\frac{n_{\qv\nu}(T)+1-f_{\kv+\qv}(T)}{\omega-(\varepsilon_{\kv+\qv}-\varepsilon_\text{F})-\omega_{\qv\nu}+i\eta}
\bigg],
\label{eq:sigma_full}
\end{equation}
\end{widetext}
where $n_{\qv\nu}(T)=[\mathrm{exp}(\omega_{\qv\nu}/T)-1]^{-1}$ and $f_{\kv}(T)=\{\mathrm{exp}[(\varepsilon_{\kv}-\varepsilon_\mathrm{F})/T]+1\}^{-1}$ are the equilibrium occupation factors for phonons and electrons, respectively. To ensure numerical stability, we introduced small imaginary terms $i\eta$ in the denominator of Eq.~(\ref{eq:sigma_full}). $\eta$ plays the role of a smearing parameter, which in our calculations is set to be equal to the temperature $T$. Denoting $\Sigma_{\kv}(\omega,T) \equiv \Sigma'_{\kv}(\omega,T) + i\Sigma''_{\kv}(\omega,T)$, the electron spectral function is then given by
\begin{equation}
A_{\kv}(\omega,T)=\frac{1}{\pi}\frac{
        |\Sigma''_{\kv}(\omega,T)|}
{
        |\omega-(\varepsilon_{\kv}-\varepsilon_\text{F})-\Sigma'_{\kv}(\omega,T)|^{2}+|\Sigma''_{\kv}(\omega,T)|^2
}.
\end{equation}

Without vertex corrections, optical conductivity for zero momentum transfer in the presence of electron-phonon coupling takes the form:
\begin{widetext}
\begin{equation}
    \mathrm{Re}[\sigma_{\alpha\beta}(0, \omega')]=\frac{2\pi e^2\hbar}{S}\sum_{\kv}\int d\omega \, \frac{f_{\kv}(\omega)-f_{\kv}(\omega+\omega')}{\omega'}v^{\alpha}_{\kv}A_{\kv}(\omega)v^{\beta}_{\kv}A_{\kv}(\omega+\omega'),
\end{equation}
\end{widetext}
where $S$ is the unit cell area.

\end{document}